\documentclass[a4paper,11pt]{article}
\usepackage{pos}
\usepackage{simpler-wick}

\title{Partially connected contributions to baryon masses in QCD+QED}

\author[a]{Anian Altherr}
\author[b]{Isabel Campos}
\author[c,d]{Alessandro Cotellucci}
\author[a]{Roman Gruber}
\author[a]{Tim Harris}
\author[a]{Javad Komijani}
\author[c,e]{Jens Lücke}
\author[a]{Marina Krstić Marinković}
\author[a]{Letizia Parato}
\author[c,e]{Agostino Patella}
\author*[b,c]{Sara Rosso}
\author[a]{Paola Tavella}

\affiliation[a]{Institut für Theoretische Physik, ETH Zürich, Wolfgang-Pauli-Str. 27, 8093 Zürich, Switzerland}
\affiliation[b]{Instituto de Física de Cantabria (IFCA) and Consejo Superior de Investigaciones Cientificas (CSIC), Avda. de Los Castros s/n, 39005 Santander, Spain}
\affiliation[c]{Humboldt Universität zu Berlin, Institut für Physik and IRIS Adlershof, Zum Großen Windkanal 6, 12489
Berlin, Germany}
\affiliation[d]{Jülich Supercomputing Centre, Forschungszentrum Jülich, D-52428 Jülich, Germany}
\affiliation[e]{DESY, Platanenallee 6, D-15738 Zeuthen, Germany}

\emailAdd{rosso@ifca.unican.es}

\abstract{Full QCD+QED simulations allow to evaluate isospin breaking corrections to hadron masses. With the \texttt{openQxD} code, we are able to perform these simulations employing C-periodic boundary conditions, implemented through a doubling of the physical lattice along one spatial direction.\\
The use of these boundary conditions introduces non-zero Wick contractions between two quark or two antiquark fields, that, in the case of the computation of baryon masses, lead to partially connected additional contributions that we expect to vanish in the infinite volume limit. These contributions are challenging because they involve an all-to-all propagator connecting one point in the physical lattice and one in the mirror lattice. We present a way to compute these corrections to the $\Omega^-$ baryon mass using a combination of point and stochastic source inversions.\\
This work is part of the program of the RC* collaboration.}

\FullConference{The 41st International Symposium on Lattice Field Theory (LATTICE2024)\\
 28 July - 3 August 2024\\
Liverpool, UK\\}

\begin{document}
\maketitle

\section{Motivation}
In order to shed light on possible violations of the Standard Model, lattice QCD simulations aim to reach percent or sub-percent precision level. To attain such precision, isospin-breaking effects must be taken into account in measurements of numerous hadronic observables, including the meson and baryon masses. These effects do not come only from the difference in mass of the light quarks (strong isospin-breaking) but also from the difference in their electric charges (electromagnetic isospin-breaking), for this reason the coupling of QCD to electrodynamics (QED) has to be investigated.\\ 
One example of the effect of these corrections is the proton-neutron mass difference, which is entirely due to a combination of strong and electromagnetic isospin-breaking effects. This work focuses precisely on the impact of these corrections on baryon masses, in particular for the $\Omega^-$, a $3/2$ spin baryon with three strange valence quarks widely used to set the scale of lattice QCD simulations (\cite{BMW}, \cite{ukqcd}, \cite{scale.set}). Our implementation of QED in finite volume, QED$_C$, makes it possible to have charged states due to a particular choice of boundary conditions in space, C-periodic, which at the same time introduce some peculiar finite-volume effects that are the subject of this study. We expect these corrections to vanish in the infinite-volume limit, since they are only an effect of the boundary conditions, and we want to quantify their magnitude with respect to the computations of baryon masses already performed by our collaboration which neglect these effects (\cite{first.res}).\\
\section{C-periodic boundary conditions and Wick contractions}
\begin{figure}
    \centering
    \includegraphics[width=0.6\linewidth]{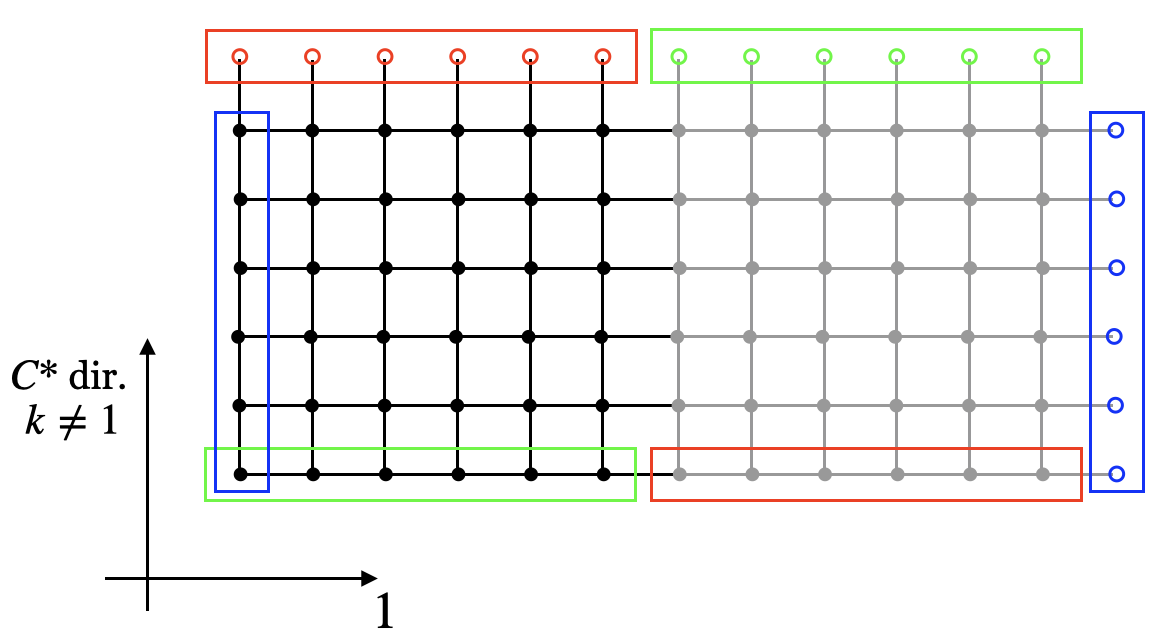}
    \caption{2-dimensional section of a 4-dimensional lattice representing the orbifold construction used to implement C-periodic boundary conditions. Figure taken from \cite{open.code}}.
    \label{fig:bound}
\end{figure}

Periodic boundary conditions in space do not allow the presence of charged states on the lattice, this is clear if we look at Gauss law to compute the total electrical charge on the lattice $Q$ starting form the charge density $\rho(x)$:
\begin{equation}
    Q = \int dx^3 \rho(x) = \int dx^3 \;\nabla \cdot E(x) = 0
    \label{eq:gauss}
\end{equation}
where the result of the integral is the difference of the electric field $E$ at the boundaries that is always zero due to its periodicity.\\
Of course, this poses a challenge for lattice QCD+QED simulations, which the RC* collaboration addresses through the use of C-periodic boundary conditions in space in the \texttt{openQ*D} code \cite{openqxd}. These boundary conditions consist in the charge conjugation of both quark and fermion fields when crossing the boundaries of the lattice:
\begin{equation}
    \begin{split}
    &A_\mu \left( x + \hat{L}_i \right) = A^C_\mu (x) = -A_\mu (x) \\
    &U(x + \hat{L}_i, \rho) = U^C(x, \rho)= U(x, \rho)^*\\
    &\psi_f \left( x + \hat{L}_i \right) = \psi_f^C (x) = C^{-1} \bar{\psi}_f^T (x)\\
    &\bar{\psi}_f \left( x + \hat{L}_i \right) = \bar{\psi}_f^C (x) = -\psi_f^T (x) C
\end{split}
\label{eq:Cbound}
\end{equation}
where $C=i\gamma_0\gamma_2$ is the charge conjugation matrix, $L_i$ is the extension of the lattice in direction $i$, $A_\mu(x)$ is the U(1) gauge field, $U(x, \rho)$ is the SU(3) gauge field and $\psi_f(x)$ is the fermion field of the quark of flavour $f$.\\
These boundary conditions allow equation (\ref{eq:gauss})  to have a non zero result because the operation of charge conjugation changes the sign of the electric field, which now becomes anti-periodic at the boundaries.\\
The boundary conditions are implemented with an orbifold construction, represented in figure (\ref{fig:bound}). In the figure is a 2-dimensional section of a 4-dimensional lattice with C-periodic boundary conditions in at least two space directions, on the horizontal axis is the first space direction and on the vertical axis another different space direction. The portion of the lattice in black represents the physical lattice, which is extended along the first space direction with an exact replica coloured in gray, the mirror lattice, where fermion and gauge fields are the charge conjugates of the ones in the physical one. Due to the fact that the square of the charge conjugation operator is the identity, in the extended lattice (physical plus mirror), we also have a periodicity of the fields that acquire the same value in the regions represented with the same colour in figure (\ref{fig:bound}).\\
This choice of boundary conditions changes the result for the Wick contractions of the quark fields, that become \cite{char.had}:
\begin{subequations}
\begin{equation}
\Big\langle \wick{\c1{q}^a_A(x) \, \c1{\overline q}^b_B(y)} \Big\rangle = D^{-1}(x, y)_{AB}^{ab}  
\end{equation}
\begin{equation}
\Big\langle \wick{\c1{q}^a_A(x) \, \c1{q}^{b,T}_{B}(y)} \Big\rangle = - D^{-1}(x, y +\hat{L_i})_{AB}^{ad} \, C_{db}  
\label{eq:wickq}
\end{equation}
\begin{equation}
\Big\langle \wick{\c1{\overline q}^{A,T}_{a}(x) \, \c1{\overline q}^b_B(y)} \Big\rangle = C^{ad} \, D^{-1}(x + \hat{L_i}, y)_{db}^{AB}
\label{eq:wick.qbar}
\end{equation}
\end{subequations}
where $a,b,d$ are Dirac indices and $A,B$ are colour indices.
We now have a non zero result for the contraction of two quark and two antiquark fields. 
In next section we will show how these new contributions are the reason for partially disconnected diagrams in the two-point function of the $\Omega^-$ baryon.
\section{Contributions to the two-point function}
Measuring hadron masses on the lattice consists in measuring two-point correlation functions of interpolating operators with the quantum numbers needed to have a good superposition with the desired hadron states. To compute the $\Omega^-$ mass we can define the following interpolating operators:
\begin{subequations}
    \begin{equation}
        v^{m;d}(x) = \sum_{\substack{abc\\ABC}} W^{d;m}_{abc;ABC} \psi^C_c(x) \psi^A_a(x) \psi^B_b(x)
    \end{equation}
    \begin{equation}
        \overline{v}^{m;d}(x) = \sum_{\substack{abc\\ABC}} \overline W^{d;m}_{abc;ABC} \overline{\psi}^B_b(x) \overline{\psi}^A_a(x) \overline{\psi}^C_c(x)
    \end{equation}
\end{subequations}
where:
\begin{subequations}
    \begin{equation}
        W^{d;m}_{abc;ABC} = \sum_{l=1}^3 \epsilon^{ABC} [ P^{ml}_{dc} \Gamma^l_{ab} + P^{ml}_{db} \Gamma^l_{ac} + P^{ml}_{da} \Gamma^l_{cb} ]
        \label{eq:w}
    \end{equation}
    \begin{equation}
         \overline W^{d;m}_{abc;ABC} = \sum_{l=1}^3 \epsilon^{ABC} [ P^{ml}_{cd} \Gamma^l_{ab} + P^{ml}_{bd} \Gamma^l_{ac} + P^{ml}_{ad} \Gamma^l_{cb} ]
    \label{eq:wbar}
    \end{equation}
    \begin{equation}
        P^{ml} = [ \delta^{ml} Id_{4\times4} - \frac{1}{3} \gamma^{m} \gamma^{l} ] 
    \end{equation}
    \begin{equation}
        \Gamma^{l} = C\gamma^{l}
    \end{equation}
\end{subequations}
the $\psi$ field is the strange quark field, lowercase latin letters $a,b,c,d$ are Dirac indices, lowercase latin letters $m,l$ are spin indices, capital latin letters are colour indices, 
 $P^{ml}$ is the 3/2 spin projector, $\epsilon$ is the total antisymmetric tensor. The three terms summed in equations (\ref{eq:w}) and (\ref{eq:wbar}) are introduced to take into account the symmetry the interpolating operators need to have in the exchanges between the quark fields, since they all have the same flavour.\\
 With these interpolating operators we can construct the two-point correlation function projected to positive parity:
\begin{equation}
    C(x_0-y_0) = \Big\langle \sum_{\substack{dd'\\m}} \sum_{\textbf{x}\textbf{y}} P^+_{d'd} \;v^{m;d}(x) \overline{v}^{m;d'}(y) \Big\rangle
    \label{eq:2.point}
\end{equation}
where $P^+$ is the positive parity projector defined as:
\begin{equation}
    P^+ = \frac{1+\gamma^0}{2}
\end{equation}
To compute the result of equation (\ref{eq:2.point}), we need to consider all the possible contractions of the quark and antiquark fields. We have a group of contributions that we also obtain in the case of periodic boundary conditions in space. These arise from all the possible Wick contractions of a quark and an antiquark field, so that the two space-time points are connected by three quark propagators, for this reason we call these contributions three-quark connected. \\
The two-point function in this case is:
\begin{equation}
    \begin{split}
        C(x_0-y_0)= \Big\langle \sum_{\textbf{x}\textbf{y}} \sum_{\substack{d'd\\m}}  \sum_{\substack{abc\\a'b'c'}}\sum_{\substack{\\ABC\\A'B'C'}} \overline W^{d';m}_{a'b'c';A'B'C'}  P^+_{d'd}  W^{d;m}_{abc;ABC} \wick{\c1{\psi}^A_a(x) \c1{\overline{\psi}}^{A'}_{a'}(y)} \, \wick{\c1{\psi}^B_b(x) \c1{\overline{\psi}}^{B'}_{b'}(y)} \\  \,\wick{\c1{\psi}^C_c(x)\c1{\overline{\psi}}^{C'}_{c'}(y)}  \Big\rangle
    \end{split}
\end{equation}
Nevertheless, according to equations (\ref{eq:wickq}) and (\ref{eq:wick.qbar}) Wick contractions between two quark fields and two antiquark fields are non-zero with C-periodic boundary conditions. This makes it possible to have another kind of contributions where two quark and two antiquark fields defined at the same space-time point are contracted between each other, in this case the two points are connected by only one quark propagator. These contributions, which we call one-quark connected, are the subject of the present study.  The two-point function in this case is:
\begin{equation}
    \begin{split}
        C(x_0-y_0)= \Big\langle \sum_{\textbf{x}} \sum_{\substack{d'd\\m}}  \sum_{\substack{\alpha bc\\\alpha'b'c'}}\sum_{\substack{\\ABC\\A'B'C'}} \overline W^{d';m}_{a'b'c';A'B'C'}  P^+_{d'd}  W^{d;m}_{abc;ABC} \wick{\c1{\overline{\psi}}^{A'}_{a'}(0) \c1{\overline{\psi}}^{B'}_{b'}(0)} \wick{  \c1 \psi_{c}^{C}(x)\c1{\overline{\psi}}^{C'}_{c'}(0) } \\\wick{\c1{\psi}^B_b(x)\c1{\psi}^A_a(x) } \Big\rangle
    \end{split}
\end{equation}
We can substitute the results in equations (\ref{eq:wickq}) and (\ref{eq:wick.qbar}), where a translation of $L_i$ along any $\hat{i}$ space direction moves the fields to the mirror lattice, for simplicity we choose the first space direction $\hat{1}$ and fix its extension to $L_1=L$. Substituting we have:
\begin{equation}
    \begin{split}
        C(x_0-y_0)= - \Big\langle \sum_{\textbf{x}\textbf{y}} \sum_{\substack{a \alpha bc\\a'\alpha'b'c'\\ }}\sum_{\substack{\\ABC\\A'B'C'}} T_{a' b' c',a b c}^{A'B'C',ABC} C_{a'\alpha'} D^{-1}(y+L\hat{1},y)^{A'B'}_{\alpha' b'} D^{-1}(x,y)^{CC'}_{cc'} \\D^{-1}(x,x+L\hat{1})^{BA}_{b\alpha}  C_{\alpha a} \Big\rangle
    \end{split}
    \label{eq:2.p.fin}
\end{equation}
where we defined the tensor 
\begin{equation}
    T_{a' b' c',a b c} = \sum_{\substack{d'd\\ml'l}} \Gamma^{l'}_{a'b'}  P^{ml'}_{c'd'}  P^+_{d'd} P^{ml}_{dc} \Gamma^l_{ba} 
    \label{eq:T}
\end{equation}
\section{Strategy for computing one-quark connected contributions}
To compute two-point functions of interpolating operators on the lattice one crucial element is the inversion of the Dirac operator. This operator is a large sparse matrix defined on the entire lattice with two Dirac and two colour indices and its inversion is usually a very computationally demanding section of the simulations. The one-quark connected contributions we want to compute are characterized by the presence of inverses of the Dirac operator computed between one space-time point and its translation to the mirror lattice. This means that, when considering the inversion at the sink location $x$, we need to compute the matrix for all the points of the lattice. The strategy we adopted is to use stochastic sources to perform this inversion and point sources for the inversion at point $y$.\\
Regarding the stochastic source inversion, we are interested in employing the identity relation for stochastic sources:
\begin{equation}
    \frac{1}{N_s} \sum_n \chi(x)^{(n)\dagger A}_a \chi(y) ^{(n)B}_b = \delta_{AB}\delta_{ab}\delta_{xy}
\end{equation}
To get:
\begin{equation}
    D^{-1}(x;x+L\hat{1})^{BA}_{b\alpha} = \frac{1}{N_s} \sum_n \left[D^{-1}\chi^{(n)}\right]_{b}^{B}(x)  \;\; \chi^{\dagger(n)}(x+L\hat{1})^{A}_{\alpha}
\end{equation}
While for the inversion of the Dirac operator using point sources, we can define a point source located at point $y$ as:
\begin{equation}
    \eta(z')^{(B'b';y)}_{Vv}  = \delta_{VB'} \delta_{v b'} \delta_{z'_0,y_0} \delta_{\textbf{z'} \textbf{y}}
\end{equation}
Substituting the definition of the sources in equation (\ref{eq:2.p.fin}) for the two-point function we get:
\begin{equation}
    \begin{split}
       &C(x_0-y_0) = -\Big\langle \sum_{\textbf{x}\textbf{y}} \sum_{\substack{a\alpha bc\\a'\alpha'b'c'd'e'}}\sum_{\substack{\\ABC\\A'B'C'D'E'}} T_{a' b' c',a b c}^{A'B'C',ABC} C_{a'\alpha'} D^{-1}(y+L\hat{1},z)^{A'D'}_{\alpha'd'}\\ & \eta^{(B'b';y)}(z)^{D'}_{d'} D^{-1}(x,w)_{ce'}^{CE'} \eta^{(C'c';y)}(w)^{E'}_{e'}  \frac{1}{N_s} \sum_n \left[D^{-1}\chi^{(n)}\right]_{b}^{B}(x)  \;\; \chi^{\dagger(n)}(x+L\hat{1})^{A}_{\alpha} C_{\alpha a} \Big\rangle
    \end{split}
    \label{eq:csources}
\end{equation}
Rewriting the result of the point inversions for simplicity as:
\begin{equation}
    \xi(z)^{(B' b';y)}_{A'\alpha'}  = \sum_{\substack{vVz'}} D^{-1}(z;z')^{A'V}_{\alpha'v} \eta(z')^{(B'b';y)}_{Vv}
\end{equation}
we finally have:
\begin{equation}
    \begin{split}
       C(x_0-y_0) = -\Big\langle \sum_{\textbf{x}\textbf{y}} \sum_{\substack{a\alpha bc\\a'\alpha'b'c'}}\sum_{\substack{\\ABC\\A'B'C'}} T_{a' b' c',a b c}^{A'B'C',ABC} C_{a'\alpha'} \xi(y+L\hat{1})^{(B'b';y)}_{A'\alpha'}  \xi(x)^{(C'c';y)}_{Cc} \\ \frac{1}{N_s} \sum_n \left[D^{-1}\chi^{(n)}\right]_{b}^{B}(x)  \;\; \chi^{\dagger(n)}(x+L\hat{1})^{A}_{\alpha} C_{\alpha a} \Big\rangle
    \end{split}
\end{equation}
This quantity is what we have implemented in the code.
\section{Status and future plans}
To check consistency, we are testing the code verifying that the desired invariance properties are displayed by the result. At the moment we are working on testing the invariance under gauge transformations. This check consists in applying a random gauge transformation to the gauge fields and the sources and verifying that we get the same result as without the gauge transformation, up to machine precision. To understand the needed transformations of the sources we can look at equation (\ref{eq:csources}) under a transformation of the gauge fields, these enter in the computation of the Dirac operator and their transformation imply that it transforms as:
\begin{equation}
    D(x,y)^{AB}_{ab} \rightarrow \sum_{CD} G(x)^{AC} D(x,y)^{CD}_{ab} G^{\dagger}(y)^{DB}
\end{equation}
We can easily see that this imply that the stochastic and point sources need to transform as:
\begin{subequations}
    \begin{equation}
        \chi(v)^D \rightarrow \sum_{F} G(v)^{DF} \chi(v)^F
    \end{equation}
    \begin{equation}
        \eta^{(Bb',y)}(z)_{Vv} \rightarrow \sum_{FH} G(z)^{VF} \eta^{(Hb',y)}(z)_{Fv} G^{\dagger}(y)^{HB'} = \eta^{(Bb',y)}(z)_{Vv}
    \end{equation}
\end{subequations}
where in the last equation we used the definition of the point source.\\
Having completed the test of gauge invariance, we will test invariance under translations in space, where we will check that the results computed with and without a space translation are equal up to machine precision.
After the testing phase, we plan to perform measurements of one-quark connected contributions on several ensembles generated by our collaboration with C-periodic boundary conditions. In particular, we want to extend the measurements of the baryon masses where only three-quark connected contributions are considered, as in \cite{first.res} and other computations that we are performing at the moment on three ensembles not included in that work.\\
The aim of this study is being able to compare the size of one-quark connected corrections and the bulk of the computation. We will check whether these corrections vanish in the infinite-volume limit and determine whether they need to be included in computations at the finite volumes we are considering.


\begin{thebibliography}{99}

\bibitem{BMW} S. Borsanyi, S. Durr, Z. Fodor, C. Hoelbling, S. D. Katz, S. Krieg, L. Lellouch, T. Lippert, A. Portelli, K. K. Szabo, and B. C. Toth, \emph{Ab initio calculation of the neutron-proton mass difference}, \href{https://doi.org/10.1126/science.1257050}{Science \textbf{347} (2015) 1452}  [\href{https://doi.org/10.48550/arXiv.1406.4088}{\texttt{1406.4088}}]

\bibitem{ukqcd} T. Blum, P.A. Boyle, N.H. Christ, J. Frison, N. Garron, R.J. Hudspith, T. Izubuchi, T. Janowski, C. Jung, A. Jüttner, C. Kelly, R.D. Kenway, C. Lehner, M. Marinkovic, R.D. Mawhinney, G. McGlynn, D.J. Murphy, S. Ohta,  A. Portelli,  C.T. Sachrajda, and A. Soni, \emph{Domain wall QCD with physical quark masses}, \href{https://doi.org/10.1103/PhysRevD.93.074505}{Phys. Rev. D \textbf{93} (2016) 074505}  [\href{https://doi.org/10.48550/arXiv.1411.7017}{\texttt{1411.7017}}]

\bibitem{scale.set} N. Miller, L. Carpenter, E. Berkowitz, C. Cheng Chang, B. Hörz, D. Howarth, H. Monge-Camacho,, E. Rinaldi, D. A. Brantley, C. Körber, C. Bouchard, M.A. Clark, A. Singh Gambhir, C. J. Monahan, A. Nicholson, P. Vranas and A. Walker-Loud, \emph{Scale setting the Möbius domain wall fermion on gradient-flowed HISQ action using the omega baryon mass and the gradient-flow scales $t_0$ and $w_0$} \href{https://doi.org/10.1103/PhysRevD.103.054511}{Phys. Rev. D \textbf{103} (2021) 054511} [\href{https://doi.org/10.48550/arXiv.2011.12166}{\texttt{2011.12166}}]

\bibitem{first.res} L. Bushnaq, I. Campos, M. Catillo, A. Cotellucci, M. Dale, P. Fritzsch, J. Lücke, M. Marinković, A. Patella, N. Tantalo, \emph{First results on QCD+QED with C* boundary conditions}, \href{https://doi.org/10.1007/JHEP03(2023)012}{JHEP \textbf{03} (2023) 012 }[\href{https://doi.org/10.48550/arXiv.2209.13183}{\texttt{2209.13183}}]

\bibitem{openqxd} RC* collaboration, \href{https://gitlab.com/rcstar/openQxD}{\texttt{openQ*D} code}

\bibitem{open.code} I. Campos, P. Fritzsch, M. Hansen, M. Marinković, A. Patella, A. Ramos, N. Tantalo, \emph{openQ*D code: a versatile tool for QCD+QED simulations}, \href{https://doi.org/10.1140/epjc/s10052-020-7617-3}{Eur. Phys. J C \textbf{80} (2020) 195}  [\href{https://doi.org/10.48550/arXiv.1908.11673}{\texttt{1908.11673}}]

\bibitem{char.had} B. Lucini, A. Patella, A. Ramos, N. Tantalo, \emph{Charged hadrons in local finite-volume QED+QCD with C* boundary conditions},  \href{https://doi.org/10.1007/JHEP02(2016)076}{JHEP \textbf{02} (2016) 076 } [\href{https://doi.org/10.48550/arXiv.1509.01636}{\texttt{1509.01636}}]

\end{thebibliography}
\end{document}